\documentclass[12pt]{article}
\usepackage{amsmath}
\usepackage{breqn}
\usepackage{cite}

\begin{document}

\title{\bf Investigation of generalised uncertainty principle effects on FRW cosmology}

\author{{\"{O}zg\"{u}r \"{O}kc\"{u}$^{a,b}$\hspace{1mm}} \\
$^a${\small {\em Department of
	Physics, Faculty of Science, Istanbul University,
		}}\\{\small {\em Istanbul, 34134, Türkiye}} \\
  $^b${\small {\em Theoretical and Computational Physics Research Laboratory, Istanbul University,
		}}\\{\small {\em Istanbul, 34134, Türkiye}} }
\maketitle

\begin{abstract}
Based on the entropy-area relation from Nouicer's generalised uncertainty
principle (GUP), we derive the GUP modified Friedmann equations from the
first law of thermodynamics at apparent horizon. We find a minimum apparent
horizon due to the minimal length notion of GUP. We show that the energy
density of universe has a maximum and finite value at the minimum apparent
horizon. Both minimum apparent horizon and maximum energy density imply
the absence of the Big Bang singularity. Moreover, we investigate the GUP
effects on the deceleration parameter for flat case. Finally, we examine
the validity of generalised second law (GSL) of thermodynamics. We show
that GSL always holds in a region enclosed by apparent horizon for the
GUP effects. We also investigate the GSL in $\Lambda CDM$ cosmology and
find that the total entropy change of universe has a maximum value in the
presence of GUP effects. To better grasp the effects of Nouicer's GUP on
cosmology, we compare our results with those obtained from quadratic GUP
(QGUP).
\end{abstract}

\section{Introduction}
\label{introduction}

Thermodynamical aspects of gravity have been an attractive research field
in theoretical physics since the discovery of black hole thermodynamics
\cite{Bekenstein1972,Bekenstein1973,Bardeen1973,Hawking1974,Bekenstein1974,Hawking1975}.
Black hole as a thermodynamic system has entropy and temperature proportional
to its horizon area and surface gravity, respectively. Considering a black
hole as a thermodynamic system transforms an absolute absorber into a perfect
laboratory to investigate the deep connection between the gravitation,
quantum mechanics, and thermodynamics. Based on the notion of black hole
thermodynamics, Jacobson obtained the Einstein field equation from thermodynamical
arguments \cite{Jacobson1995}. Employing the entropy-area relation with
Clausius relation $\delta Q=TdS$, he derived the field equation as an equation
of state. Here, $\delta Q$, $T$ and $dS$ are energy flux, Unruh temperature
and the change in entropy, respectively. After the pioneering work of Jacobson,
there have been many studies targeted the thermodynamical aspects of gravity
in the literature
\cite{Padmanabhan2002,Eling2006,Paranjape2006,Kothawala2007,Padmanabhan2007,Cai2005,Akbar2006,Akbar2007,Cai2007a,Cai2007b,Cai2008,Sheykhi2010a,Awad2014,Salah2017,Kouwn2018,Okcu2020,Luo2023,Alsabbagh2023,Sheykhi2010b,Sheykhi2007a,Sheykhi2007b,Sheykhi2009,Sheykhi2018,Nojiri2019,Lymperis2018,Karami2011,Sheykhi2019,Saridakis2020,Barrow2021,Saridakis2021,Anagnostopoulos2020,Sheykhi2021,Asghari2021,Sheykhi2022,Sheykhi2022b,Asghari2022,Sheykhi2023,Sheykhi2023b,Lymperis2021,Drepanou2022,Odintsov2023,Abreu2022,Coker2023,Genarro2022,Odintsov2023b,Nojiri2022,Nojiri2022b,Nojiri2023,Nojiri2023b}.
The studies aimed to understand obtaining Einstein field equation from
the first law of thermodynamics can be found in Refs.
\cite{Padmanabhan2002,Eling2006,Paranjape2006,Kothawala2007,Padmanabhan2007}.
Inspired by Jacobson's paper, $(n+1)$-dimensional Friedmann equations were
obtained from the first law of thermodynamics, $-dE=T_{h}dS_{h}$, at apparent
horizon by Cai and Kim \cite{Cai2005}. Here $-dE$ is the energy flux crossing
the apparent horizon for the infinitesimal time interval at fixed horizon
radius. The temperature and entropy of the apparent horizon are given by
\cite{Cai2005}
\begin{equation}
\label{Temp-entropy1}
T_{h}=\frac{1}{2\pi \tilde{r_{A}}},\qquad \qquad S_{h}=
\frac{A_{h}}{4},
\end{equation}
where $A_{h}$ and $\tilde{r_{A}}$ correspond to area and apparent horizon,
respectively.\footnote{We adopt the units $\hbar =c=G_{N}=L^{2}_{Pl}=1$ throughout the
paper.} Furthermore, in the rest of their seminal paper
\cite{Cai2005}, they also derived the Friedmann equations from the entropy-area
relations of Gauss-Bonnet gravity and Lovelock gravity theories, where
the standard entropy-area relation is break down. In Ref.~\cite{Akbar2006}, Akbar and Cai obtained the Friedmann equations in scalar-tensor
and $f(R)$ gravity theories by following the arguments of Ref.~\cite{Cai2005}. Although the Friedmann equations can successfully be obtained
in Eq.~(\ref{Temp-entropy1}), the temperature is an approximation for a
fixed apparent horizon. Thus the temperature is not proportional to the
surface gravity at apparent horizon. Moreover, the equation of state is
only limited with the vacuum energy or de Sitter spacetime for this approximation.
The surface gravity of apparent horizon is given by
\cite{Cai2005,Hayward1998}
\begin{equation}
\label{kappa}
\kappa =-\frac{1}{\tilde{r_{A}}}\left (1-
\frac{\dot{\tilde{r_{A}}}}{2H\tilde{r_{A}}}\right ),
\end{equation}
where dot denotes the derivative with respect to time and $H$ corresponds
to Hubble parameter. Assuming temperature is proportional to surface gravity
\cite{Akbar2007}
\begin{equation}
\label{Temp}
T_{h}=\frac{\kappa}{2\pi}=-\frac{1}{2\pi \tilde{r_{A}}}\left (1-
\frac{\dot{\tilde{r_{A}}}}{2H\tilde{r_{A}}}\right ),
\end{equation}
with the standard entropy-area relation, Akbar and Cai showed that the
first law of thermodynamics at apparent horizon defined as
\begin{equation}
\label{firstLaw}
dE=T_{h}dS_{h}+WdV,
\end{equation}
where $E=\rho V$ and $W$ correspond to the total energy in volume
$V$ enclosed by the apparent horizon and the work density, respectively.
Then, many papers devoted to Friedmann equations and apparent horizon thermodynamics
have been intensively studied in the Refs.
\cite{Cai2007a,Cai2007b,Cai2008,Sheykhi2010a,Awad2014,Salah2017,Kouwn2018,Okcu2020,Luo2023,Alsabbagh2023,Sheykhi2010b,Sheykhi2007a,Sheykhi2007b,Sheykhi2009,Sheykhi2018,Nojiri2019,Lymperis2018,Karami2011,Sheykhi2019,Saridakis2020,Barrow2021,Saridakis2021,Anagnostopoulos2020,Sheykhi2021,Asghari2021,Sheykhi2022,Sheykhi2022b,Asghari2022,Sheykhi2023,Sheykhi2023b,Lymperis2021,Drepanou2022,Odintsov2023,Abreu2022,Coker2023,Genarro2022,Odintsov2023b,Nojiri2022,Nojiri2022b,Nojiri2023,Nojiri2023b}.

It is widely known that the standard entropy-area relation is not valid
and should be corrected in the case of various theories. For example, it
is well-known that logarithmic correction to black hole entropy arises
due to quantum gravity effects at the Planck scale
\cite{Govindarajan2001,Mann1998,Sen2013}. The various quantum gravity approaches
to the modifications of Friedmann equations such as loop quantum gravity
\cite{Cai2008,Sheykhi2010a}, modified Heisenberg principle
\cite{Awad2014,Salah2017,Kouwn2018,Okcu2020,Luo2023,Alsabbagh2023} were
investigated in the literature. Other interesting modifications of entropy-area
relation are considered in the context of generalised statistics such as
Tsallis statistics \cite{Tsallis2013} and Kaniadakis statistics
\cite{Kaniadakis2002,Kaniadakis2005}. Moreover, inspired by COVID-19, Barrow
proposed that the area of horizon may be deformed in the context of quantum
gravity effects and entropy is the power law function of its area
\cite{Barrow2020}. Recently, the effects of fractional quantum mechanics
were investigated on black hole thermodynamics in Ref.~\cite{Jalalzadeh2021}. Interestingly, the fractional entropy
\cite{Jalalzadeh2021} is similar to Tsallis \cite{Tsallis2013} and Barrow
\cite{Barrow2020} entropies although all entropies have different origins.
The extensions of Refs.
\cite{Tsallis2013,Kaniadakis2002,Kaniadakis2005,Barrow2020,Jalalzadeh2021}
to cosmological cases, namely modified Friedmann equations can be found
for Tsallis entropy in Refs.
\cite{Sheykhi2018,Nojiri2019,Lymperis2018}, Kaniadakis entropy in Refs.
\cite{Lymperis2021,Drepanou2022}, Barrow entropy in Refs.
\cite{ Saridakis2020,Barrow2021,Saridakis2021,Anagnostopoulos2020,Sheykhi2021,Asghari2021,Sheykhi2022,Genarro2022}
and fractional entropy in Ref.~\cite{Coker2023}.

Modifications of standard uncertainty principle may provide defining the
physics both at Planck scale and large distance scales. GUP takes into
account the momentum uncertainty correction while the extended uncertainty
principle (EUP) takes into account the position uncertainty correction.
GUP predicts a minimum measurable length at Planck scale while EUP may
predict a minimum measurable momentum. Taking into account effects of GUP
and EUP leads to third kind of modified uncertainty principle known as
generalized and extended uncertainty principle (GEUP). It predicts the
notions of minimum measurable length and minimum measurable momentum. Numerous
models of modified uncertainty principle were suggested in Refs.
\cite{Maggiore1993,Scardigli1999,Kempf1995,Bambi2008,Nozari2012,Filho2016,Chung2018,Lake2019,Chung2019,Dabrowski2019,Lake2020,Lake2021,Mureika2019,Du2022,Segreto2023,Ali2009,Ali2011,Vagenas2019,Pedram2012,Nouicer2006}.
Both GUP and EUP provide new insights on black thermodynamics
\cite{Nouicer2006,Adler2001,Medved2004,Nozari2008,Arraut2009,Banerjee2010x,Nozari2012b,Ali2012,Feng2016,Xiang2009,Scardigli2020a,Hassanabadi2021,Lutfuoglu2021,Sun2018,Ma2018,Ong2018,Okcu2020b,Bolen2005,Han2008,MoradpourEUP2019,HassanabadiEUP2019,ChungEUP2019,HamilEUP2021,HamilEUP2021b,ChenEUP2019,Park2007,Majumder2013,Chen2013,Anacleto2015,Okcu2022}.
For example, the black holes are prevented from total evaporation in the
GUP case and a black hole remnant occurs at the final stage
\cite{Adler2001}. On the other hand, EUP implies a minimum temperature
for black holes \cite{HassanabadiEUP2019}. Both GUP and EUP have been widely
studied in cosmological applications
\cite{Awad2014,Salah2017,Kouwn2018,Okcu2020,Luo2023,Alsabbagh2023,Scardigli2011,Nenmeli2021,Das2022,Barca2022,Barca2023,Barca2023b}.
Extensions of modified uncertainty principle to cosmological scenario lead
to modified Friedmann equations
\cite{Awad2014,Salah2017,Kouwn2018,Okcu2020,Luo2023,Alsabbagh2023}. Based
on the simplest form of GUP, the modified Friedmann equations were derived
in Ref.~\cite{Awad2014}. They found an upper bound energy density of universe
at minimal length. In Ref.~\cite{Salah2017}, a cyclic universe model was
defined from the GUP modified Friedmann equations. In Ref.~\cite{Okcu2020}, we obtained the modified Friedmann equations from a GUP
model obtained in doubly special relativity \cite{Chung2018}. Similarly,
we obtained a maximum and finite energy density due to minimal length at
Planck scale. We also investigated the effects of GUP on the deceleration
parameter. The authors of Ref.~\cite{Luo2023} investigated the baryon asymmetry
for the EUP modified Friedmann equations. They also obtained the constraints
on EUP parameter from observations. Modified Friedmann equations in GEUP
case can be found in Ref.~\cite{Kouwn2018} where the author obtained the
bounds for GEUP parameters from observations. Recently, GEUP corrected
Friedmann equations were investigated in Ref.~\cite{Alsabbagh2023}. The
authors studied the deceleration parameters and showed that GEUP may be
an alternative to dark energy at late time expansion.\footnote{For a recent
review on GUP, the reader may refer to Ref.~\cite{Bosso2023}.}

In this paper, our aim is to obtain the GUP modified Friedmann equations
from the first law of thermodynamics at apparent horizon. The modified
Friedmann equations reveal that the initial singularity is removed since
GUP implies the minimal measurable length at Planck scale. Moreover, a
detailed investigation on GSL is crucial to understand the GUP effects.
Especially, we would like to understand how GUP affects the GSL in the
$\Lambda CDM$ cosmology. In order to modify the Friedmann equations, we
use Nouicer's GUP \cite{Nouicer2006} whose details are given in the next
section. Nouicer proposed the following higher-order GUP, which is consistent
with QGUP (\ref{QGUP}) up to the leading order
\begin{equation}
\label{NouicerGUP1}
\Delta X\Delta P\geq \frac{1}{2}e^{\alpha ^{2}\Delta P^{2}}.
\end{equation}
Including the higher-order terms in GUP generate a quantitative correction
to physics at Planck scale. Especially, higher-order terms make a significant
contribution to black hole thermodynamics \cite{Nouicer2006}. Besides,
understanding the effects of higher-order terms is also crucial for cosmology
at Planck scale. Therefore, we compare the results obtained from Nouicer's
GUP with those obtained from QGUP.

The paper is organised as follows. In Section~\ref{SectNGUPEntropy}, we
start a brief review on Nouicer's GUP. Then, we obtained the modified entropy-area
relation from Nouicer's GUP. In Section~\ref{SectModiFried}, using modified
entropy-area relation, we obtain the Friedmann equations. In Section~\ref{SectDecelPara}, GUP effects on deceleration parameter are investigated.
In Section~\ref{SectGSL}, we check the validity of GSL and explore the
GUP effects on GSL in the $\Lambda CDM$ cosmology. Finally, we discuss
our results in Section~\ref{Conlc}. The details of QGUP modified Friedmann
equations are given in Appendix~\ref{Appendix}.

\section{Noucier's GUP and entropy$-$area relation}
\label{SectNGUPEntropy}

In this section, we first begin to review Nouicer's GUP
\cite{Nouicer2006}. Then, we obtain the modified entropy-area relation
based on the methods of Xiang and Wen \cite{Xiang2009}. We start to consider
the nonlinear relation $p=f(k)$ between the momentum $p$ and the wave vector
$k$ of particle \cite{Hossenfelder2006,Hossenfelder2006b}. This relation
must satisfy the following conditions:
\begin{itemize}
\item The relation $p=f(k)$ reduces to usual relation $p=k$ for the lower
energies than the Planck energy.
\item The relation $p=f(k)$ approaches to maximum value at Planck scale
for the higher energies.
\end{itemize}
Following the above conditions, Nouicer proposed the modified position
and momentum operators
\begin{equation}
\label{operators}
X=ie^{\alpha ^{2}P^{2}}\frac{\partial}{\partial p}, \quad \quad P=p,
\end{equation}
where $\alpha $ is a dimensionless positive GUP parameter. These operators
lead to modified commutator
\begin{equation}
\label{commutator}
\left [X,P\right ]=ie^{\alpha ^{2}P^{2}}.
\end{equation}
Using the relations
$\left \langle P^{2n}\right \rangle \geq \left \langle P^{2}\right
\rangle ^{n}$,
$\left (\Delta P\right )^{2}=\left \langle P^{2}\right \rangle -
\left \langle P\right \rangle ^{2} $ with the above equation, one can obtain
the GUP in Eq.~(\ref{NouicerGUP1}) for the physical state
$\left \langle P\right \rangle =0$. The square of Eq.~(\ref{NouicerGUP1})
can be written by
\begin{equation}
\label{Lambert}
W(u)e^{W(u)}=u,
\end{equation}
where $W$ is the Lambert function \cite{Corless1996} and we define
$W(u)=-2\alpha ^{2}\Delta P^{2}$ and
$u=-\frac{\alpha ^{2}}{2\Delta X^{2}}$. For $0\geq u\geq -1/e$, using Eq.~(\ref{NouicerGUP1}), the momentum uncertainty is given
\begin{equation}
\label{momentumUncertainty}
\Delta P=\frac{1}{2\Delta X}\exp \left \{ -\frac{1}{2}W\left (-
\frac{1}{e}\frac{\Delta X_{0}^{2}}{\Delta X^{2}}\right )\right \},
\end{equation}
where $\Delta X_{0}$ is the minimum position uncertainty and is given by
\begin{equation}
\label{minimumPositionUnceratinty}
\Delta X_{0}=\sqrt{\frac{e}{2}}\alpha .
\end{equation}
It is obtained from the condition $u\geq -1/e$.

Following the arguments of Ref.~\cite{Xiang2009}, let us start to obtain
the modified entropy-area relation. When a black hole absorbs a particle,
the change in area is defined by \cite{Bekenstein1973}
\begin{equation}
\label{ChangeOfArea}
\Delta A\sim bm,
\end{equation}
where $b$ and $m$ are the particle's size and mass. We consider two limitations
on particle's size and mass. The particle is defined by a wave packet in
quantum mechanics. We have $b\sim \Delta X$ since the width of the wave
packet is defined by the particle's size. The second limitation comes from
the fact that the momentum uncertainty is not bigger than the particle
size. Thus we have $m\geq \Delta P$. With these two limitations, we can
write
\begin{equation}
\label{ChangeOfArea2}
\Delta A\sim bm\geq \Delta X\Delta P.
\end{equation}
Using Eqs.~(\ref{momentumUncertainty}), (\ref{minimumPositionUnceratinty})
and $\Delta X\sim 2r_{h}$ event horizon, the change in area is given by
\begin{equation}
\label{ChangeOfArea3}
\Delta A\geq \frac{\gamma}{2}\exp \left \{ -\frac{1}{2}W\left (-
\frac{1}{8}\frac{\alpha ^{2}}{r_{h}^{2}}\right )\right \},
\end{equation}
where $\gamma $ is a calibration factor. Using above expression with minimum
increase of entropy, $(\Delta S)_{min}=\ln 2$, the GUP modified entropy-area
relation is given by
\begin{equation}
\label{EntropyAreaRelation}
\frac{dS_{h}}{dA}=\frac{\Delta S_{min}}{\Delta A_{min}}=\frac{1}{4}
\exp \left \{ \frac{1}{2}W\left (\frac{-\alpha ^{2}}{8r_{h}^{2}}
\right )\right \},
\end{equation}
where we find $\gamma =8\ln 2$ since the above equation must give
$dS_{h}/dA=1/4$ in the limit $\alpha \rightarrow 0$. In the next section,
we use Eq.~(\ref{EntropyAreaRelation}) to modify the Friedmann equations.

\section{Modified Friedmann equations from the first law of thermodynamics}
\label{SectModiFried}

We first begin a concise review on the basic elements of Friedmann-Robertson-Walker
(FRW) universe. The line element of FRW universe is defined by
\cite{Cai2005}
\begin{equation}
\label{lineElement}
ds^{2}=h_{ab}dx^{a}dx^{b}+\tilde{r}^{2}d\Omega ^{2},
\end{equation}
where $\tilde{r}=a(t)r$, $a(t)$ is the scale factor, $x^{a}=(t,r)$, and
$h_{ab}=diag\left (-1,a^{2}/(1-kr^{2})\right )$ is the two-dimensional
metric. $k=$ $-1$, $0$, and $1$ present to the open, flat, and closed universe,
respectively. The expression of apparent horizon $\tilde{r_{A}}$ is given
by \cite{Cai2005}
\begin{equation}
\label{apparentHor}
\tilde{r_{A}}=ar=\frac{1}{\sqrt{H^{2}+k/a^{2}}},
\end{equation}
where the Hubble parameter is defined by $H=\dot{a}/a$. In the following,
we consider the matter and energy of universe as a perfect fluid. So we
have
\begin{equation}
\label{energyMomentumTensor}
T_{\mu \nu}=(\rho +p)u_{\mu}u_{\nu}+pg_{\mu \nu},
\end{equation}
where $\rho $, $p$ and $u_{\mu}$ correspond to energy density, pressure
and four-velocity of the fluid. The conservation of energy-momentum
tensor ($\nabla _{\mu}T^{\mu \nu}=0$) gives the continuity equation as
\begin{equation}
\label{continuityEqu}
\dot{\rho}+3H(\rho +p)=0.
\end{equation}
The work by the volume change of universe is defined by work density
\cite{Hayward1998}
\begin{equation}
\label{workDensity}
W=-\frac{1}{2}T^{ab}h_{ab}=\frac{1}{2}(\rho -p),
\end{equation}
and the corresponding volume is given by \cite{Akbar2007}
\begin{equation}
V=\frac{4}{3}\pi \tilde{r_{A}}^{3}.
\label{volume}
\end{equation}
Employing the Eqs.~(\ref{continuityEqu}) and (\ref{volume}), the differential
of the total energy of universe is given by
\begin{equation}
\label{dE}
dE=\rho dV+Vd\rho =4\pi \rho \tilde{r_{A}}^{2}d\tilde{r_{A}}-4\pi (
\rho +p)\tilde{r_{A}}^{3}Hdt.
\end{equation}
From Eqs.~(\ref{workDensity}) and (\ref{volume}), we can obtain
\begin{equation}
\label{WdV}
WdV=2\pi (\rho -p)\tilde{r_{A}}^{2}d\tilde{r_{A}}.
\end{equation}
Since entropy is defined by the area, $S=S(A)$, one can give the general
expressions of entropy as follows \cite{Awad2014}:
\begin{equation}
\label{f(A)Rel}
S_{h}=\frac{f(A_{h})}{4},
\end{equation}
and its differential can be written by
\begin{equation}
\label{diffOfEntropy}
\frac{dS_{h}}{dA_{h}}=\frac{f'(A_{h})}{4},
\end{equation}
where prime denotes derivative with respect to the apparent horizon area
$A_{h}=4\pi \tilde{r_{A}}^{2}$. Comparing Eq.~(\ref{EntropyAreaRelation})
with the above equation, we find
\begin{equation}
\label{diffOfEntropy2}
f'(A_{h})=\exp \left \{ \frac{1}{2}W\left (
\frac{-\alpha ^{2}}{8\tilde{r_{A}}^{2}}\right )\right \} .
\end{equation}
Finally, we find $T_{h}dS_{h}$ as
\begin{equation}
\label{TdS}
T_{h}dS_{h}=-\left (1-\frac{\dot{\tilde{r_{A}}}}{2H\tilde{r_{A}}}
\right )f'(A_{h})d\tilde{r_{A}}.
\end{equation}
Now, we can obtain the modified Friedmann equations since we have all the
necessary ingredients. Substituting Eqs.~(\ref{dE})-(\ref{TdS}) into Eq.~(\ref{firstLaw}) and using the differential form of apparent horizon
\begin{equation}
\label{difAppa}
d\tilde{r_{A}}=-H\tilde{r_{A}}^{3}\left (\dot{H}-\frac{k}{a^{2}}
\right )dt,
\end{equation}
we find
\begin{equation}
\label{dynamicalEq}
4\pi (\rho +p)\tilde{r_{A}}^{3}Hdt=f'(A_{h})d\tilde{r_{A}}.
\end{equation}
Employing the continuity equation (\ref{continuityEqu}) with the above
equation, the differential form of Friedmann equation is given by
\begin{equation}
\label{firstEquation}
\frac{f'(A_{h})}{\tilde{r_{A}}^{3}}d\tilde{r_{A}}=-\frac{4\pi}{3}d
\rho .
\end{equation}
Integrating the above equation with Eq.~(\ref{diffOfEntropy2}) gives the
first Friedmann equation
\begin{equation}
\label{firstFriedmannEquaNGUPWithCons}
\frac{8}{9\alpha ^{2}}e^{\frac{3}{2}W\left (
\frac{-\alpha ^{2}}{8\tilde{r_{A}}^{2}}\right )}\left (1+3W\left (
\frac{-\alpha ^{2}}{8\tilde{r_{A}}^{2}}\right )\right )+C=-
\frac{4\pi \rho}{3},
\end{equation}
in the limit $\tilde{r_{A}}\rightarrow \infty $, this equation must reduce
to standard Friedmann equation, i.e.,
$H^{2}+\frac{k}{a^{2}}=\frac{8\pi}{3}\rho +\frac{\Lambda}{3}$, thus we
set the integration constant
$C=-\frac{8}{9\alpha ^{2}}+\frac{\Lambda}{6}$. So we can write the first
Friedmann equation as
\begin{equation}
\label{firstFriedmannEquationa}
\frac{8}{9\alpha ^{2}}\left [e^{\frac{3}{2}W\left (
\frac{-\alpha ^{2}}{8\tilde{r_{A}}^{2}}\right )}\left (1+3W\left (
\frac{-\alpha ^{2}}{8\tilde{r_{A}}^{2}}\right )\right )-1\right ]=-
\frac{4\pi}{3}\rho -\frac{\Lambda}{6}.
\end{equation}
Finally, combining Eqs.~(\ref{diffOfEntropy2}) and (\ref{difAppa}) with
Eq.~(\ref{dynamicalEq}), the dynamical equation can be obtained
\begin{equation}
\label{secondEquation}
e^{\frac{1}{2}W\left (\frac{-\alpha ^{2}}{8\tilde{r_{A}}^{2}}\right )}
\left (\dot{H}-\frac{k}{a^{2}}\right )=-4\pi (\rho +p).
\end{equation}
These equations reduce to standard Friedmann equations in the limit
$\alpha \rightarrow 0$.

The striking feature of the first Friedmann equation (\ref{firstFriedmannEquationa})
comes from the argument of Lambert function, i.e., the condition
$\frac{\alpha ^{2}}{8\tilde{r_{A}}^{2}}\leq \frac{1}{e}$ gives the minimal
apparent horizon
\begin{equation}
\label{rAMin}
\tilde{r_{A}}^{min}=\sqrt{\frac{e}{8}}\alpha .
\end{equation}
This minimal apparent horizon appears due to minimal length notion of GUP
\cite{Awad2014,Salah2017,Okcu2020,Alsabbagh2023}. It implies that the singularity
is removed at the beginning of the Universe. Moreover, the energy density
does not diverge anymore since the minimum apparent horizon has a finite
value. Using Eq.~(\ref{rAMin}) in the first Friedmann equation (\ref{firstFriedmannEquationa})
and neglecting the cosmological constant, the maximum energy density is
given by
\begin{equation}
\label{maxRho}
\rho ^{max}=\frac{2\left (2+e^{3/2}\right )}{3\pi e^{3/2}\alpha ^{2}}.
\end{equation}
It is clear that both $\tilde{r_{A}}^{min}$ and $\rho ^{max}$ recover the
standard results in the limit $\alpha \rightarrow 0$, i.e.,
$\tilde{r_{A}}^{min}$ approaches zero while $\rho ^{max}$ diverges. Using
Eq.~(\ref{apparentHor}), the Friedmann equations can be expressed in term
of Hubble parameter
\begin{eqnarray}
\label{Fried2}
-\frac{4\pi}{3}\rho -\frac{\Lambda}{6}=\frac{8}{9\alpha ^{2}}\left [e^{
\frac{3}{2}W\left (\frac{-\alpha ^{2}}{8}\left (H^{2}+\frac{k}{a^{2}}
\right )\right )}\left (1+3W\left (\frac{-\alpha ^{2}}{8}\left (H^{2}+
\frac{k}{a^{2}}\right )\right )\right )-1\right ],
\nonumber
\\
-4\pi (\rho +p)=e^{\frac{1}{2}W\left (\frac{-\alpha ^{2}}{8}\left (H^{2}+
\frac{k}{a^{2}}\right )\right )}\left (\dot{H}-\frac{k}{a^{2}}\right ).
\end{eqnarray}
When the GUP effects are tiny, these equations can be expanded in powers
of $\alpha $
\begin{eqnarray}
\label{Fried2Expanded}
\frac{8\pi}{3}\rho +\frac{\Lambda}{3}=\left (H^{2}+\frac{k}{a^{2}}
\right )-\frac{1}{32}\left (H^{2}+\frac{k}{a^{2}}\right )^{2}\alpha ^{2}+...,
\nonumber
\\
-4\pi (\rho +p)=\left (\dot{H}-\frac{k}{a^{2}}\right )\left [1-
\frac{1}{16}\left (H^{2}+\frac{k}{a^{2}}\right )\alpha ^{2}+...
\right ].
\end{eqnarray}
Neglecting the GUP correction, one can easily recover the usual Friedmann
equations.

In order to clearly reveal the effects of Nouicer's GUP, we compare the
Nouicer's GUP modified results with the results corrected by QGUP. For
the QGUP case, we give minimum apparent horizon and maximum energy density
in Eq.~(\ref{QGUPMinRAAndMaxRho}). In Table~\ref{TableA}, we show
$\tilde{r_{A}}^{min}$ and $\rho _{max}$ for the different values of
$\alpha $ in Nouicer's GUP case. In Table~\ref{TableB}, we present the
same quantities for the different values of $\alpha $ in QGUP case. Nouicer's
GUP implies that the Universe starts at the minimum apparent horizon bigger
than one obtained from QGUP case while maximum energy density obtained
from Nouicer's GUP is smaller than one obtained from QGUP. From Tables~\ref{TableA} and \ref{TableB}, one can see that
$\tilde{r_{A}}^{min}$ increases and $\rho _{max}$ decreases when
$\alpha $ increases.

\begin{table}
\centering
\begin{tabular}{llllll}
\hline
 & $\alpha =0.6$ & $\alpha =0.8$ & $\alpha =1$ & $\alpha =1.2$ & $\alpha =1.4$\tabularnewline
\hline
$\tilde{r_{A}}^{min}$ & $0.35$ & $0.466$ & $0.583$ & $0.7$ & $0.816$\tabularnewline
$\rho _{max}$ & $0.853$ & $0.48$ & $0.307$ & $0.213$ & $0.157$\tabularnewline
$H_{max}$ & $1.734$ & $1.301$ & $1.041$ & $0.867$ & $0.743$\tabularnewline
\hline
\end{tabular}
\caption{$\tilde{r_{A}}^{min}$, $\rho _{max}$ and $H_{max}$ for the different
values of $\alpha $ in Nouicer's GUP case.}
\label{TableA}
\end{table}
\begin{table}
\centering
\begin{tabular}{llllll}
\hline
 & $\alpha =0.6$ & $\alpha =0.8$ & $\alpha =1$ & $\alpha =1.2$ & $\alpha =1.4$\tabularnewline
\hline
$\tilde{r_{A}}^{min}$ & $0.3$ & $0.4$ & $0.5$ & $0.6$ & $0.7$\tabularnewline
$\rho _{max}$ & $1.105$ & $0.622$ & $0.398$ & $0.276$ & $0.203$\tabularnewline
$H_{max}$ & $3.333$ & $2.5$ & $2$ & $1.667$ & $1.429$\tabularnewline
\hline
\end{tabular}
\caption{$\tilde{r_{A}}^{min}$, $\rho _{max}$ and $H_{max}$ for the different
values of $\alpha $ in QGUP case.}
\label{TableB}
\end{table}

\section{Deceleration parameter}
\label{SectDecelPara}

We would like to analyse the effects of GUP on the deceleration parameter.
It is defined by
\begin{equation}
\label{decelPara}
q=-1-\frac{\dot{H}}{H^{2}}.
\end{equation}
The positivity of $q$ means the decelerated expansion while the negativity
of $q$ means the accelerated expansion. We choose the equation of state
as $p=\omega \rho $. We restrict our analysis with the flat case
$k=0$ since it is consistent with cosmological observations
\cite{Odintsov2011}. Combining Friedmann equations in (\ref{Fried2}) with
the deceleration parameter, we can obtain
\begin{equation}
\label{decelPara2}
q=-1-
\frac{4(1+\omega )}{H^{2}e^{\frac{1}{2}W\left (\frac{-\alpha ^{2}H^{2}}{8}\right )}}
\left \{ \frac{2}{3\alpha ^{2}}\left [e^{\frac{3}{2}W\left (
\frac{-\alpha ^{2}H^{2}}{8}\right )}\left (1+3W\left (
\frac{-\alpha ^{2}H^{2}}{8}\right )\right )-1\right ]+
\frac{\Lambda}{8}\right \}.
\end{equation}
Now, we can investigate the deceleration parameter at the beginning of
the Universe. Remembering the arguments of Lambert function, i.e., the
condition $\frac{\alpha ^{2}H^{2}}{8}\leq \frac{1}{e}$ yields the maximum
\begin{equation}
\label{HMAX}
H_{max}=\frac{2\sqrt{2}}{e\alpha},
\end{equation}
at the initial stage. For the maximum value of Hubble parameter, the deceleration
parameter at the initial stage is given by
\begin{equation}
\label{decelParaAtHMax}
q(H_{max})=\omega +(1+\omega )\left (
\frac{\exp \left \{ 2-\frac{1}{2}W\left (-\frac{1}{e^{2}}\right )\right \} }{3}+
\frac{1}{3W\left (-\frac{1}{e^{2}}\right )}\right ),
\end{equation}
or equivalently
\begin{equation}
\label{decelParaAtHMax2}
q(H_{max})\approx \omega +0.564(1+\omega ).
\end{equation}
Equation of state parameter $\omega $ must satisfy the condition
$\omega <-0.361$ for the acceleration at the inflationary stage. Since
GUP effects are negligible for the radiation and matter dominated eras,
the deceleration parameter can be expanded
\begin{equation}
\label{decelParaSer}
q=\frac{1+3\omega}{2}-\frac{(1+\omega )\Lambda}{2H^{2}}+
\frac{3H^{2}(1+\omega )\alpha ^{2}}{64}-
\frac{1(1+\omega )\Lambda \alpha ^{2}}{32}+...
\end{equation}
Thus, the deceleration parameters for radiation ($\omega =1/3$) and matter
($\omega =0$) dominated eras can be expressed as
\begin{equation}
q_{rad}=1+\frac{H^{2}\alpha ^{2}}{16},\qquad q_{m}=\frac{1}{2}+
\frac{3H^{2}\alpha ^{2}}{64},
\label{decelParMat}
\end{equation}
respectively. The results imply that the universe is more decelerated for
the radiation and matter dominated eras when the GUP effects are considered.

Now, let us compare our results with the results obtained from QGUP. In
Eqs.~(\ref{decelQGUP}), we give the QGUP corrected deceleration parameter.
In Table~\ref{TableA} and Table~\ref{TableB}, we present $H_{max}$ at the
initial stage for Nouicer's GUP and QGUP, respectively. In the Nouicer's
GUP case, $H_{max}$ at the initial stage is smaller than $H_{max}$ obtained
from QGUP case. From Tables~\ref{TableA} and \ref{TableB}, one can see
that $H_{max}$ decreases while $\alpha $ increases.

\section{Generalised second law}
\label{SectGSL}

In this section, we want to confirm the validity of GSL, which states the
total entropy of matter fields and horizon cannot decrease with time, for
the GUP effects. We start to reorganise Eq.~(\ref{dynamicalEq}) as follows:
\begin{equation}
\label{intEq2}
\dot{r_{A}}=4\pi (\rho +p)H\tilde{r_{A}}^{3}e^{-\frac{1}{2}W\left (
\frac{-\alpha ^{2}}{8\tilde{r_{A}}^{2}}\right )}.
\end{equation}
From Eqs.~(\ref{diffOfEntropy2}), (\ref{TdS}) and (\ref{intEq2}), we can
find
\begin{equation}
\label{GSL}
T_{h}\dot{S_{h}}=4\pi (\rho +p)H\tilde{r_{A}}^{3}\left [1-2\pi (\rho +p)
\tilde{r_{A}}^{2}e^{-\frac{1}{2}W\left (
\frac{-\alpha ^{2}}{8\tilde{r_{A}}^{2}}\right )}\right ].
\end{equation}
The second law at apparent horizon may be violated for the accelerated
expansion phase. Therefore, we must also consider the entropy of matter
field inside the horizon, i.e., we must check the GSL. The Gibbs equation
is given by \cite{Izquierdo2006}
\begin{equation}
\label{GE}
T_{f}dS_{f}=d(\rho V)+pdV=Vd\rho +(\rho +p)dV,
\end{equation}
where $T_{f}$ and $S_{f}$ are the temperature and entropy of matter fields,
respectively. In order to avoid nonequilibrium thermodynamics and spontaneous
energy flow between horizon and matter, thermal equilibrium condition ($T_{h}=T_{f}$)
is assumed. Otherwise, the deformation of FRW geometry is unavoidable
\cite{Izquierdo2006}. From Eqs.~(\ref{continuityEqu}), (\ref{volume}),
(\ref{intEq2}) and (\ref{GE}), the change in entropy of matter fields can
be expressed by
\begin{equation}
\label{GSL2}
T_{h}\dot{S_{f}}=-4\pi (\rho +p)H\tilde{r_{A}}^{3}\left (1-4\pi (
\rho +p)\tilde{r_{A}}^{2}e^{-\frac{1}{2}W\left (
\frac{-\alpha ^{2}}{8\tilde{r_{A}}^{2}}\right )}\right ).
\end{equation}
At last, combining Eqs.~(\ref{GSL}) and (\ref{GSL2}), the total entropy
evolution is written by
\begin{equation}
\label{GSLT}
T_{h}(\dot{S_{h}}+\dot{S_{f})}=8\pi ^{2}(\rho +p)^{2}H\tilde{r_{A}}^{5}e^{-
\frac{1}{2}W\left (\frac{-\alpha ^{2}}{8\tilde{r_{A}}^{2}}\right )}.
\end{equation}
The right hand side of the above expression never decreases with respect
to time. Therefore, we conclude that the GSL is always satisfied for all
eras of the universe for any spatial curvature.

Now, in order to understand how GUP affects the total evolution of entropy,
we focused on a more specific case, namely the $\Lambda CDM$ scenario.
We first begin to solve the first Friedmann equation (\ref{Fried2}). So
we find
\begin{equation}
\label{LCDMH}
H^{2}=
\frac{8\exp \left \{ \frac{1}{3}\left (2W\left (\frac{\sqrt{e}}{4}\left (2-3\pi \alpha ^{2}\rho \right )\right )-1\right )\right \} \left (1-2W\left (\frac{\sqrt{e}}{4}\left (2-3\pi \alpha ^{2}\rho \right )\right )\right )}{3\alpha ^{2}},
\end{equation}
for $k=0$. Let us simplify the above equation. The total energy density
is defined by
\begin{equation}
\label{totalEnergyDensity}
\rho =\rho _{m}+\rho _{r}+\rho _{\Lambda},
\end{equation}
where the matter density $\rho _{m}$, the radiation density
$\rho _{r}$ and the cosmological constant energy density
$\rho _{\Lambda}$ have the dependencies
$\rho _{m}=\frac{\rho _{m0}}{a^{3}}$,
$\rho _{r}=\frac{\rho _{r0}}{a^{4}}$ and
$\rho _{\Lambda}=\rho _{\Lambda 0}$ \cite{Ryden}. Zero subscript denotes
the current value. On the other hand, the matter, radiation and cosmological
constant density parameters are defined by
$\Omega _{m0}=\frac{\rho _{m0}}{\rho _{c0}}, \Omega _{r0}=
\frac{\rho _{r0}}{\rho _{c0}}, \Omega _{\Lambda 0}=1-\Omega _{m0}-
\Omega _{r0}$, respectively. Here, the current critical density
$\rho _{c0}$ is given by $\rho _{c0}=\frac{3H_{0}^{2}}{8\pi}$ and
$H_{0}$ is the current Hubble parameter. Using the above definitions with
redshift parameter $1+z=\frac{a_{0}}{a}$ and taking the present scale factor
$a_{0}=1$, one can write
\begin{equation}
\label{LCDMRho}
\frac{\rho}{\rho _{c0}}=\frac{\Omega _{m0}}{a^{3}}+
\frac{\Omega _{r0}}{a^{4}}+\Omega _{\Lambda 0}=\Omega _{m0}(1+z)^{3}+
\Omega _{r0}(1+z)^{4}+\Omega _{\Lambda 0}.
\end{equation}
\begin{figure}
\centering
\includegraphics[width=11cm]{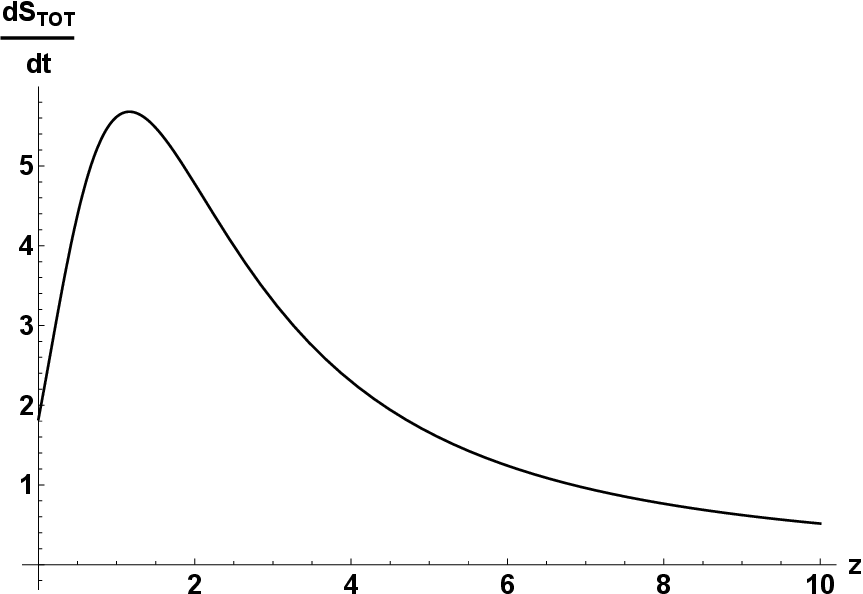}
\caption{Total entropy change versus the redshift parameter. We take $\Omega
_{m}=0.3$, $\Omega _{r}=0.0001$, $H_{0}=1$ and $\alpha =0$.}
\label{totEntClass}%
\end{figure}
\begin{figure}
\centering 
\includegraphics[width=11cm]{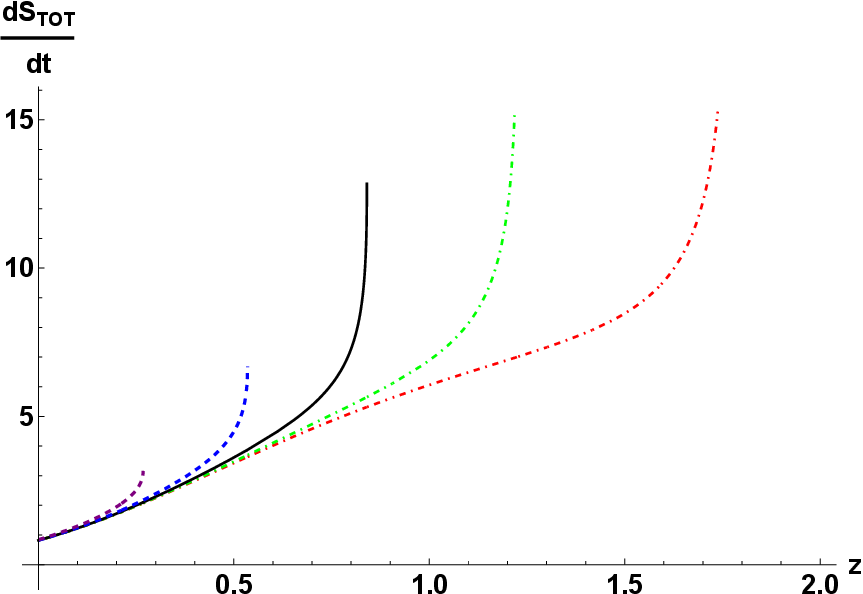}
\caption{Total entropy change versus the redshift parameter. From top to bottom,
the curves correspond to $\alpha =1.4, 1.2, 1, 0.8, 0.6$ for Nouicer's GUP. We
take $\Omega _{m}=0.3$, $\Omega _{r}=0.0001$ and $H_{0}=1$.}
\label{totGUPEnt}%
\end{figure}
\begin{figure}
\centering 
\includegraphics[width=11cm]{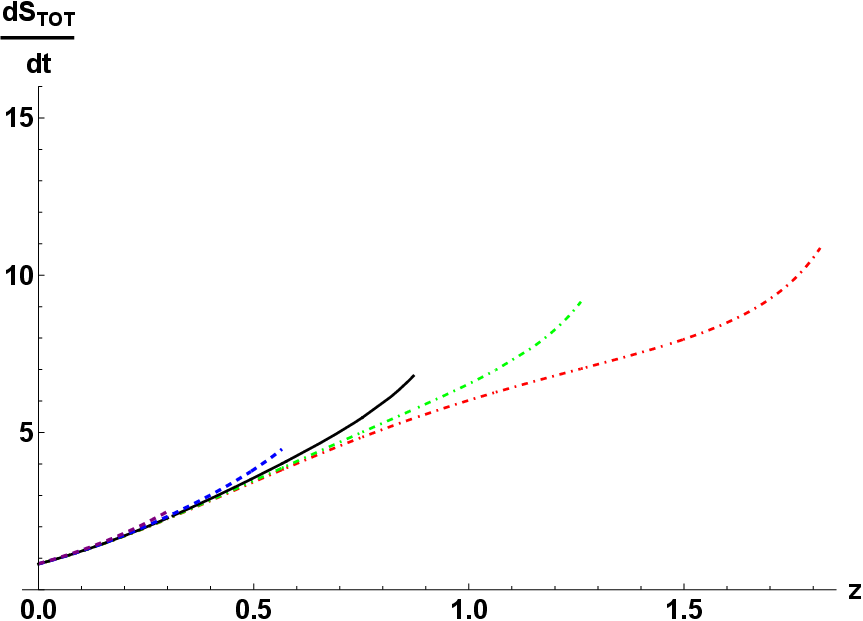}	
\caption{Total entropy change versus the redshift parameter. From top to bottom,
the curves correspond to $\alpha =1.4, 1.2, 1, 0.8, 0.6$ for QGUP. We take
$\Omega _{m}=0.3$, $\Omega _{r}=0.0001$ and $H_{0}=1$.}
\label{totQGUPEnt}%
\end{figure}
Using Eq.~(\ref{LCDMRho}) in the arguments of Lambert function (\ref{LCDMH}),
we get
\begin{equation}
\label{LCDMu}
x=\frac{\sqrt{e}}{4}\left (2-3\pi \alpha ^{2}\rho \right )=
\frac{\sqrt{e}}{4}\left (2-\frac{9H_{0}^{2}\alpha ^{2}}{8}\left [
\Omega _{r0}(1+z)^{4}+\Omega _{m0}(1+z)^{3}+\Omega _{\Lambda 0}
\right ]\right ).
\end{equation}
Finally, we can express the GUP modified Hubble function for the
$\Lambda CDM$ cosmology as follows:
\begin{equation}
\label{LCDMHz}
H_{\Lambda CDM}(z)=
\frac{2\sqrt{2}e^{\frac{2W(x)-1}{6}}\sqrt{1-2W(x)}}{\sqrt{3}\alpha}
\end{equation}
Substituting Eqs.~(\ref{Temp}), (\ref{apparentHor}) and the second Friedmann
equation (\ref{Fried2}) into Eq.~(\ref{GSLT}), we find
\begin{equation}
\label{LCDMTotEnt1}
\dot{S_{h}}+\dot{S_{f}}=\pi \dot{H}^{2}H^{-5}\left (1+
\frac{\dot{H}}{2H^{2}}\right )^{-1}e^{\frac{1}{2}W\left (
\frac{-\alpha ^{2}H^{2}}{8}\right )}
\end{equation}
Finally, using $\dot{H}=-(1+z)H(dH/dz)$, we obtain
\begin{equation}
\label{LCDMTotEnt}
\dot{S_{h}}+\dot{S_{f}}=\pi (1+z)^{2}H^{-3}\left (\frac{dH}{dz}
\right )^{2}\left (1-\frac{(1+z)}{2H}\frac{dH}{dz}\right )^{-1}e^{
\frac{1}{2}W\left (\frac{-\alpha ^{2}H^{2}}{8}\right )}.
\end{equation}

Combining Eq.~(\ref{LCDMHz}) with the above equation, we plotted total
entropy change with respect to redshift. In Fig.~\ref{totEntClass}, we
plot the evolution of total entropy for the standard case. In Fig.~\ref{totGUPEnt}, the total entropy change is represented for the different
values of GUP parameter $\alpha $. Comparing Fig.~\ref{totEntClass} with \ref{totGUPEnt} reveals the dramatic effects on the evolution of total
entropy. As can be seen in Fig.~\ref{totGUPEnt}, the redshift parameter
has an upper bound. In contrast to GUP case, the redshift parameter is
allowed to go infinity in the standard cosmology. Since $z$ reaches infinity,
the total entropy change vanishes in the standard cosmology. On the other
hand, the total entropy change has an upper bound since $z$ has a maximum
value in the presence of GUP effects.

Again, we compare our results with the results obtained from QGUP. We give
the total entropy evolution in Eqs.~(\ref{totalEntQGUP}), (\ref{totalSQGUP})
and (\ref{totalSQGUP2}). In Fig.~\ref{totQGUPEnt}, we present the total
entropy change with respect to redshift for the QGUP case. Similarly, total
entropy change and redshift parameter have maximum values in the QGUP case.
Comparing Fig.~\ref{totGUPEnt} with \ref{totQGUPEnt} reveals that Nouicer's
GUP modified total entropy evolution has a drastic increase when $z$ approaches
to $z_{max}$. In Tables~\ref{Table1} and \ref{Table2}, we numerically give
the maximum values of the redshift parameter, Hubble parameter and total
entropy change for the various values of GUP parameter $\alpha $ in Nouicer's
GUP and QGUP cases, respectively. In the Nouicer's GUP case,
$z_{max}$ and $H(z_{max})$ are smaller than $z_{max}$ and
$H(z_{max})$ obtained from QGUP. Nouicer's GUP modified total entropy change
is bigger than QGUP modified total entropy change.

Finally, we finish this section with comments on different values of GUP
parameter. From Fig.~\ref{totGUPEnt}, Fig.~\ref{totQGUPEnt}, Table~\ref{Table1} and Table~\ref{Table2}, one can see that $z_{max}$ decreases
while $\alpha $ increases. The same behaviour can also be seen for the
maximum values of $H$ and total entropy change. Maximum values of Hubble
parameter and total entropy changes decrease while $\alpha $ increases.
It is also interesting to note that Nouicer's GUP corrected total entropy
change drastically decreases for the increasing values of $\alpha$.

\begin{table}
\centering
\begin{tabular}{llllll}
\hline
 & $\alpha =0.6$ & $\alpha =0.8$ & $\alpha =1$ & $\alpha =1.2$ & $\alpha =1.4$\tabularnewline
\hline
$z_{max}$ & $1.779$ & $1.227$ & $0.840$ & $0.535$ & $0.268$\tabularnewline
$H(z_{max})$ & $2.859$ & $2.114$ & $1.716$ & $1.430$ & $1.125$\tabularnewline
$\dot{S_{h}}+\dot{S_{f}}$ & $100.173$ & $26.276$ & $12.861$ & $6.707$ & $3.185$\tabularnewline
\hline
\end{tabular}
\caption{The maximum values of redshift parameter, Hubble parameter and the
total entropy change for the different values of $\alpha $ in Nouicer's GUP
case.}
\label{Table1}
\end{table}

\begin{table}
\centering
\begin{tabular}{llllll}
\hline
 & $\alpha =0.6$ & $\alpha =0.8$ & $\alpha =1$ & $\alpha =1.2$ & $\alpha =1.4$\tabularnewline
\hline
$z_{max}$ & $1.816$ & $1.26$ & $0.871$ & $0.566$ & $0.301$\tabularnewline
$H(z_{max})$ & $2.887$ & $2.165$ & $1.732$ & $1.443$ & $1.237$\tabularnewline
$\dot{S_{h}}+\dot{S_{f}}$ & $10.862$ & $9.156$ & $6.796$ & $4.468$ & $2.498$\tabularnewline
\hline
\end{tabular}
\caption{The maximum values of redshift parameter, Hubble parameter and the
total entropy change for the different values of $\alpha $ in QGUP case.}
\label{Table2}
\end{table}

\section{Conclusions and discussions}
\label{Conlc}

In this section, using the entropy-area relation obtained from Nouicer's
GUP \cite{Nouicer2006}, we obtained the GUP modified Friedmann equations
from the first law of thermodynamics at apparent horizon
\cite{Akbar2007}. We found a minimum apparent horizon due to the minimal
length notion of GUP. We showed that the energy density of universe is
finite and maximum at the minimum apparent horizon. Then, in order to see
the effects of GUP, we computed the deceleration parameter for flat case
and the equation of state $p=\omega \rho $. We found that $\omega $ must
satisfy the condition $\omega <-0.361$ for the initial acceleration. For
the radiation and matter dominated eras, the expansion of universe is more
decelerated since the GUP effects give the positive contribution to deceleration
parameter. Moreover, we checked the validity of GSL. We showed that the
GSL is always valid for the all eras of universe in the presence of GUP
effects. At last, we consider the GSL for the specific case, i.e.,
$\Lambda CDM$ cosmology. In contrast to standard cosmology, the redshift
parameter has a finite and maximum value. GUP also affects the total entropy
change. The total entropy change has a finite and maximum value at maximum
redshift value.

To better understand the effects of Nouicer's GUP on FRW cosmology, we
compared our results with those obtained from QGUP. We showed that universe
has beginning with bigger apparent horizon and less dense energy density
in the Nouicer's GUP case.

Our results indicate that there is no Big Bang singularity due to the minimal
apparent horizon and maximum energy density. Therefore, GUP provides more
reasonable solution at the Planck scale where the classical general relativity
fails. This feature is a well-known in the literature
\cite{Awad2014,Salah2017,Okcu2020,Alsabbagh2023}. So our results are consistent
with the recent studies. Interestingly, we found the total entropy change
has a maximum value at maximum and finite redshift for $\Lambda CDM$ cosmology.
Moreover, maximum total entropy evolution has a drastic increase in Nouicer's
GUP case. In fact, the maximum and finite value of $z$ is expected since
the Big Bang singularity is removed. However, the modified uncertainty
effects on total entropy change in $\Lambda CDM$ need further investigation.
Particularly, taking into account different forms of GUP may shed light
on this case. For example, linear and quadratic GUP (LQGUP)
$[X,P]=i(1+\alpha p+\beta p^{2})$ includes a linear term in momentum
\cite{Ali2009,Ali2011,Vagenas2019}. Clearly, LQGUP is not consistent with
QGUP and Nouicer's GUP due to the linear term in momentum. Including the
linear term may affect total entropy evolution. On the other hand, a higher
order GUP in the form of $[X,P]=\frac{i}{1-\beta P^{2}}$ is consistent
with Nouicer's GUP and QGUP up to the leading order \cite{Pedram2012}.
A quantitative correction at Planck scale can be determined for the GSL.
We hope to report effects of various GUP models on GSL in future studies.

\section*{Declaration of competing interest}
The authors declare that they have no known competing financial
interests or personal relationships that could have appeared to influence
the work reported in this paper.

\section*{Data availability}
No data was used for the research described in the article.

\section*{Acknowledgments}
 The author thanks anonymous reviewer for his/her invaluable and constructive
comments, which greatly improved the quality of paper. The author thanks
Ekrem Aydiner for valuable and fruitful discussion. This work was supported by Istanbul University Post-Doctoral Research Project: MAB-2021-38032.

\appendix
\counterwithin*{equation}{section}
\renewcommand\theequation{\thesection.\arabic{equation}}
\section{QGUP modified Friedmann equations}
\label{Appendix}

In this appendix, we will review the QGUP modified Friedmann equations
\cite{Awad2014}. We will compute the deceleration parameter and investigate
the generalised second law for the QGUP modified Friedmann equations. We
consider the QGUP given by \cite{Maggiore1993,Scardigli1999}
\begin{equation}
\label{QGUP}
\Delta X\Delta P\geq \frac{1}{2}\left (1+\alpha ^{2}\Delta P^{2}
\right ).
\end{equation}

Repeating the same calculations given in Sections~\ref{SectNGUPEntropy} and \ref{SectModiFried}, one can obtain the QGUP
modified entropy-area relation and Friedmann equations
\begin{equation}
\label{QGUPEARel}
\frac{dS_{h}}{dA_{h}}=\frac{\Delta S_{min}}{\Delta A_{min}}=
\frac{\alpha ^{2}}{32\tilde{r_{A}}^{2}}\left (1-
\sqrt{1-\frac{\alpha ^{2}}{4\tilde{r_{A}}^{2}}}\right )^{-1},
\end{equation}
\begin{equation}
\label{QGUPFirstFr}
-\frac{8\pi \rho}{3}-\frac{\Lambda}{3}=\frac{2}{3\alpha ^{2}}\left
\lfloor \left (1-\frac{\alpha ^{2}}{4\tilde{r_{A}}^{2}}\right )\left (3+2
\sqrt{1-\frac{\alpha ^{2}}{4\tilde{r_{A}}^{2}}}\right )-5\right
\rfloor ,
\end{equation}
\begin{equation}
\label{QGUPSecondFr}
-4\pi (\rho +p)\left (1-
\sqrt{1-\frac{\alpha ^{2}}{4\tilde{r_{A}}^{2}}}\right )=
\frac{\alpha ^{2}}{8\tilde{r_{A}}^{2}}\left (\dot{H}-\frac{k}{a^{2}}
\right ),
\end{equation}
respectively. These equations can be expressed in term of Hubble parameter\footnote{The reader may notice that the Friedmann equations in Eq.~(\ref{QGUPFrH})
are slightly different from the equations in Ref.~\cite{Awad2014}. (See
Eqs. (5.5) and (5.6) in Ref.~\cite{Awad2014}). Replacing
$\alpha ^{2}\rightarrow \alpha /\pi $ and
$\Lambda \rightarrow -8\pi \Lambda $ in Eq.~(\ref{QGUPFrH}), one can exactly
get the same QGUP modified Friedmann equation in Ref.~\cite{Awad2014}.
The authors of Ref.~\cite{Awad2014} defined the GUP parameter as
$\beta $ while we define the GUP parameter as $\alpha ^{2}$. Moreover,
they set $\alpha =4\pi \ell _{P}\beta $ to simplify the equations. In the
limit $\tilde{r_{A}}\rightarrow \infty $, we set the integration constant
to reproduce the standard Friedmann equations. Considering vacuum energy
density $\rho _{vac}=\Lambda $ in the limit
$\tilde{r_{A}}\rightarrow \infty $, the authors of Ref.~\cite{Awad2014}
set the integration constant. However, these arguments lead to obtaining
the cosmological constant term as $-\frac{8\pi \Lambda}{3}$ instead of
$\frac{\Lambda}{3}$ in Ref.~\cite{Awad2014}.}
\begin{eqnarray}
\label{QGUPFrH}
\frac{8\pi}{3}\left (\rho +\frac{\Lambda}{8\pi}\right )=\frac{1}{2}
\left (H^{2}+\frac{k}{a^{2}}\right )+\frac{4}{3\alpha ^{2}}\left [1-
\left (1-\frac{\alpha ^{2}}{4}\left (H^{2}+\frac{k}{a^{2}}\right )
\right )^{3/2}\right ],
\nonumber
\\
-4\pi (\rho +p)=\left (\dot{H}-\frac{k}{a^{2}}\right )
\frac{\alpha ^{2}}{8}
\frac{\left (H^{2}+\frac{k}{a^{2}}\right )}{\left [1-\left (1-\frac{\alpha ^{2}}{4}\left (H^{2}+\frac{k}{a^{2}}\right )\right )^{1/2}\right ]}.
\end{eqnarray}
From Eq.~(\ref{QGUPFirstFr}), the minimum apparent horizon and maximum
energy density are given by
\begin{equation}
\label{QGUPMinRAAndMaxRho}
\tilde{r_{A}}^{min}=\frac{\alpha}{2},\qquad \qquad \rho _{max}=
\frac{5}{4\pi \alpha ^{2}},
\end{equation}
respectively. Choosing equation of state $p=\omega \rho $ and combining
the Friedmann equations (\ref{QGUPFrH}) with Eq.~(\ref{decelPara}), one
can obtain the deceleration parameter
\begin{equation}
\label{decelQGUP}
q=-1+\frac{6(1+\omega )}{H^{2}\alpha ^{2}}\left [1+
\frac{8}{3H^{2}\alpha ^{2}}\left (1-\left (1-
\frac{\alpha ^{2}H^{2}}{4}\right )^{1/2}\right )-
\frac{2\Lambda}{3H^{2}}\right ]\left [1-\left (1-
\frac{\alpha ^{2}H^{2}}{4}\right )^{1/2}\right ]
\end{equation}
for $k=0$. From the above equation, the maximum Hubble parameter is given
by
\begin{equation}
\label{HMaxQGUP}
H_{max}=\frac{2}{\alpha}.
\end{equation}
The deceleration parameter at the initial stage is given by
\begin{equation}
\label{qMaxQGUP}
q(H_{max})=-1+\frac{5(1+\omega )}{2}.
\end{equation}
$\omega $ must satisfy the condition $\omega <-0.6$ for the acceleration
at the inflationary stage. Finally, repeating the calculations defined
in Section~\ref{SectGSL}, the total entropy evolution is given by
\begin{equation}
\label{totalEntQGUP}
T_{h}\left (\dot{S_{h}}+\dot{S_{f}}\right )=
\frac{64\pi ^{2}(\rho +p)^{2}H\tilde{r_{A}}^{7}}{\alpha ^{2}}\left (1-
\sqrt{1-\frac{\alpha ^{2}}{4\tilde{r_{A}}^{2}}}\right ).
\end{equation}
We conclude that GSL is always satisfied since the above expression never
decreases. In order to investigate the QGUP effects on total entropy evolution,
we can write Eq.~(\ref{totalEntQGUP}) as follows:
\begin{equation}
\label{totalSQGUP}
\dot{S_{h}}+\dot{S_{f}}=\frac{\pi}{8}\dot{H}^{2}H^{-3}\left (1+
\frac{\dot{H}}{2H^{2}}\right )^{-1}
\frac{\alpha ^{2}}{1-\sqrt{1-\frac{\alpha ^{2}H^{2}}{4}}}.
\end{equation}
At last, using $\dot{H}=-(1+z)H(dH/dz)$, we get
\begin{equation}
\label{totalSQGUP2}
\dot{S_{h}}+\dot{S_{f}}=\frac{\pi}{8}(1+z)^{2}H^{-1}\left (
\frac{dH}{dz}\right )^{2}\left (1-\frac{(1+z)}{2H}\frac{dH}{dz}
\right )^{-1}
\frac{\alpha ^{2}}{1-\sqrt{1-\frac{\alpha ^{2}H^{2}}{4}}}.
\end{equation}
Repeating the calculations given in Section~\ref{SectGSL}, one can find
QGUP modified $H_{\Lambda CDM}(z)$ function. We do not give the exact expression
since it is too lengthy.

\end{document}